\documentclass{article}
\usepackage{amssymb}

\usepackage{graphicx}
\usepackage{amsmath}

\input{tcilatex}

\begin{document}

\title{$\mathbf{8}$-Spinor Quantum Gravity\thanks{%
This is a somewhat expanded version of an essay submitted March 31, 2002 to
the Gravitation Research Society essay contest. This work was presented on
July 24, 2002, at the Pacific Institute of Mathematical Sciences Summer
Workshop on ``Brane-World and Supersymmetry'' in Vancouver, British Columbia.%
}}
\author{Marcus S. Cohen \\
Department of Mathematical Sciences\\
New Mexico State University\\
Las Cruces, New Mexico\\
marcus@nmsu.edu}
\date{\today}
\maketitle

\begin{abstract}
Quantum gravity has been so elusive because we have tried to approach it by
two paths which can never meet: quantum mechanics and general relativity.
These contradict each other not only in superdense regimes, but also in the 
\emph{vacuum}.

We explore a straight road to quantum gravity here---the one mandated by
Clifford-algebra covariance. This bridges the gap from microscales---where
the massive Dirac propagator is a sum over \emph{null zig-zags}---to
macroscales---where we see the \emph{energy-momentum current,} $\ast T$ and
the resulting Einstein curvature, $\ast G$. For massive particles, $\ast T$
flows in the ``cosmic time'' direction, $y^{0}$---\emph{centrifugally} in an
expanding universe.

Neighboring centrifugal currents of $\ast T$ present \emph{opposite}
spacetime vorticities $\ast G$ to the boundaries of each others' worldtubes,
so they \emph{advect}---i.e. \emph{attract}, as we show here by integrating
a Spin$^{c}$-4 Lagrangian by parts in the \emph{spinfluid} regime.

This boundary integral not only explains \emph{why} stress-energy $\ast T$
is the source for gravitational curvature $\ast G$, but also gives a \emph{%
value} for the gravitational constant, $\kappa \left( x^{0}\right) $ that
depends on the current scale factor of our expanding Friedmann $3$-brane! On
the microscopic scale, \emph{quantum gravity} appears naturally as the \emph{%
statistical mechanics} of null zig-zags of massive particles in ``imaginary
time,'' $y^{0}$.
\end{abstract}

\section{The Spinfluid Flow: Dilation-Boost Current}

We \emph{derive} Einstein's field equations here by recognizing Einstein
curvature $\ast G$ and energy-momentum $\ast T$ as different expressions for
the \emph{same} flux---the ``spinfluid current'' $3$ form---and that these
expressions must match on the \emph{boundaries} $\partial B^{4}$ of the
worldtubes of massive particles. We then look at a patch of this boundary at
a microscopic scale---where the matter current is resolved into a sum over
null zig-zags.

What is essential to recognize first is that the macroscopic \emph{%
energy-momentum} current $\ast T\left( x\right) $ is the \emph{Noether
current of the spinor (matter) fields under the active local dilation/boost
flow} $\varphi ^{\alpha }\left( x\right) $ and that the \emph{dilation
current}, or \emph{energy density}, is\emph{\ centrifugally outward} in an
expanding universe, $\mathbb{M}$ \cite{im}.

Conserved currents spring from \emph{invariances}. We use Spin$^{c}$-4, the
complexification of $\left( \text{Spin-4}\right) \times U\left( 1\right) $,
as our isometry group on expanding, curved space, $\mathbb{M}\equiv \left(
T,S_{3}\left( T\right) \right) $; a family of space-like hypersurfaces
parametrized by \emph{cosmic time} $T$. After Sachs \cite{sachs}, we call $E$
the \emph{Einstein} group. Cosmic expansion and boost-covariance demand that 
$E$ be \emph{nonunitary}---i.e. have \emph{Hermitian }$\left( H\right) $, as
well as \emph{anti-Hermitian }$\left( aH\right) $ generators.

Let's view our spacetime $\mathbb{M}$ as filled with a spin$^{c}$-4 fluid or 
\emph{spinfluid:} an inhomogeneous distribution of $4$ spinor fields, $\psi
_{I}\left( x\right) $, and $4$ dual spinors $\psi ^{I}\left( x\right) $.
Suppose that each may be created from the homogeneous \emph{vacuum}
distribution $\hat{\psi}$ by active-local Einstein ($E_{A}$) transformations 
\cite{im}, \cite{xueg}: 
\begin{equation}
\begin{array}{c}
\psi _{I}\left( x\right) =\exp \left( \frac{i}{2}\zeta _{I}^{\alpha }\left(
x\right) \sigma _{\alpha }\right) \hat{\psi}_{I}\equiv g_{I}\left( x\right) 
\hat{\psi}_{I}\text{,} \\ 
\psi ^{I}\left( x\right) =\hat{\psi}^{I}\exp \left( \frac{i}{2}\zeta
_{\alpha }^{I}\left( x\right) \sigma ^{\alpha }\right) \equiv \hat{\psi}%
^{I}g^{I}\left( x\right) \text{;} \\ 
\alpha =\left( 0,1,2,3\right) \text{.}
\end{array}
\label{1}
\end{equation}

In the $PTC$-symmetric \emph{geometrical-optics} regime, the path-dependent
phase shifts $\mathbf{d}\zeta _{I}^{\alpha }\left( x\right) =-\mathbf{d}%
\zeta _{\alpha }^{I}\left( x\right) $ are \emph{complex:} 
\begin{equation}
g^{I}\mathbf{d}g_{I}\equiv g_{I}^{-1}\mathbf{d}g_{I}=\frac{i}{2}\mathbf{d}%
\zeta _{I}^{\alpha }\left( x\right) \sigma _{\alpha }=\frac{i}{2}\left[ 
\mathbf{d}\theta _{I}^{\alpha }\left( x\right) -i\mathbf{d}\varphi
_{I}^{\alpha }\right] \sigma _{\alpha }\equiv \Omega _{I}\left( x\right) 
\text{.}  \label{2}
\end{equation}
The $\Omega _{I}\left( x\right) $ are the \emph{spin connections} or \emph{%
vector potentials}---the $gl\left( 2,\mathbb{C}\right) $-valued $1$ forms
that enter into the covariant derivatives of each spinor field: 
\begin{equation*}
\nabla _{\beta }\psi _{I}\equiv \left( \partial _{\beta }+\Omega _{I\beta
}\left( x\right) \right) \psi _{I}\text{.}
\end{equation*}
The $\Omega _{I}$ record the phase shift of each spin frame \cite{penrose}, 
\cite{capp} in any direction at $x\in \mathbb{M}$ due to local sources---and
of the global (vacuum) distribution.

Their path dependences, or \emph{holonomies} 
\begin{equation}
g^{I}\mathbf{dd}g_{I}\equiv K_{I}=\mathbf{d}\Omega _{I}+\Omega _{I}\wedge
\Omega _{I}  \label{3}
\end{equation}
are the \emph{spin curvatures}---the \emph{fields}. Their \emph{%
anti-Hermitian} ($aH$) parts are the $u\left( 1\right) \times su\left(
2\right) $ (electroweak) and $su\left( 3\right) $ (strong) fields. Their 
\emph{Hermitian} ($H$) parts are the \emph{gravitational} fields; these
measure the path dependence of the local \emph{dilation-boost flow} $\mathbf{%
d}\varphi ^{\alpha }\left( x\right) $ \cite{im}, \cite{xueg}.

A concentration of \emph{mass} at Minkowsky-space position $x\equiv \left(
x^{0},x^{1},x^{2},x^{3}\right) \in \mathbb{M}$ \emph{is} a localized $4$%
-momentum current $\mathbf{d}\varsigma ^{0}\left( x\right) $ propagating in
the \emph{cosmic time} ($y^{0}$) direction---which is not directly visible
to us as dwellers in a spacelike (constant $y^{0}$) cross section.

However, $y^{0}$ enters kinematically \cite{im}, \cite{penrose} as the \emph{%
imaginary} part of a \emph{complex time} variable $z^{0}\equiv x^{0}+iy^{0}$%
. The imaginary parts $y^{j}$ of $z^{j}\equiv x^{j}+iy^{j}$ are the $3$%
-momentum density.

Now, if the phase flow 
\begin{equation}
\mathbf{d}\varsigma ^{\alpha }\left( z\right) \equiv \frac{\partial \zeta
^{\alpha }}{\partial z^{\beta }}\mathbf{d}z^{\beta }+\frac{\partial \zeta
^{\alpha }}{\partial \bar{z}^{\beta }}\mathbf{d}\bar{z}^{\beta }  \label{4}
\end{equation}
were \emph{analytic,} it would obey the Cauchy-Riemann equations: 
\begin{equation}
\frac{\partial \varsigma ^{\alpha }}{\partial \bar{z}^{\beta }}%
=0\Longrightarrow \frac{\partial \theta ^{\alpha }}{\partial x^{\beta }}=%
\frac{\partial \varphi ^{\alpha }}{\partial y^{\beta }}\text{;\quad }\frac{%
\partial \theta ^{\alpha }}{\partial y^{\beta }}=-\frac{\partial \varphi
^{\alpha }}{\partial x^{\beta }}\text{.}  \label{5}
\end{equation}
We could then detect the dilation current, or \emph{rest energy,} by the 
\emph{frequency} 
\begin{equation}
\frac{\partial \theta ^{0}}{\partial x^{0}}=\frac{\partial \varphi ^{0}}{%
\partial y^{0}}  \label{6}
\end{equation}
of the matter wave, $\psi $. But \emph{energy} is the Noether charge 
\begin{equation}
\int_{B_{3}}\frac{\partial \mathcal{L}}{\partial \left( \partial _{0}\psi
_{I}\right) }\left[ \frac{\partial \psi _{I}}{\partial x^{0}}\right] d^{3}V=%
\text{%
h\hskip-.2em\llap{\protect\rule[1.1ex]{.325em}{.1ex}}\hskip.2em%
}\int_{B_{3}}\left( \frac{\partial \theta ^{0}\left( x\right) }{\partial
x^{0}}\right) e^{1}\wedge e^{2}\wedge e^{3}  \label{7}
\end{equation}
under Minkowsky-time translation, and\emph{\ is} proportional to the
frequency (\ref{6}), which we \emph{can} detect! In quantum mechanics, the
proportionality constant is 
h\hskip-.2em\llap{\protect\rule[1.1ex]{.325em}{.1ex}}\hskip.2em%
.

But Q.M. is \emph{incompatible} with general relativity (G.R.)---not only at
small scales or high densities---but in the \emph{vacuum!} The problem is
that the divergent Q.M. vacuum energy would produce enough spacetime
curvature to roll up our space to a point!

The \emph{solution} is a fundamental theory from which both Q.M. and G.R.
derive in different regimes. We illustrate below using a \emph{Nonlinear
Multispinor} (N.M.) model. In the macroscopic limit, we recover G.R. and in
the microscopic view, the spinfluid flow resolves into a sum over null
zig-zags---the Dirac propagator \cite{penrose}, \cite{ord}.

\section{Vacuum Energy in Spinfluid Models}

Covariance of the Dirac equations in curved spacetime \cite{davies} rests on
the local \emph{spinorization maps} or \emph{Maurer-Cartan 1 forms} 
\begin{equation}
\begin{array}{c}
S\equiv q_{\alpha }\left( x\right) e^{\alpha }\left( x\right) :e_{\beta
}\left( x\right) \longrightarrow q_{\beta }\left( x\right) \text{;} \\ 
\bar{S}\equiv \bar{q}_{\alpha }\left( x\right) e^{\alpha }\left( x\right)
:e_{\beta }\left( x\right) \longrightarrow \bar{q}_{\beta }\left( x\right) 
\text{.}
\end{array}
\label{8}
\end{equation}
These assign local generators $q_{\beta }\left( x\right) \in gl\left( 2%
\mathbb{C}\right) _{L}$, $\bar{q}_{\beta }\left( x\right) \in gl\left( 2%
\mathbb{C}\right) _{R}$ of the ``internal'' Lie algebra to each spacetime
increment $e_{\alpha }\left( x\right) $. These obey the \emph{Clifford
algebra} of $\mathbb{M}$: 
\begin{equation}
\left[ q_{\alpha }\bar{q}_{\beta }+q_{\beta }\bar{q}_{\alpha }\right] \left(
x\right) =2g_{\alpha \beta }\left( x\right) \sigma _{0}\text{.}  \label{9}
\end{equation}
The overbar denotes \emph{quaternionic} conjugation (space, or $P$ reversal).

The covariant and contravariant tetrads are sums of \emph{null tetrads:}
tensor products of some fundamental $L$ and $R$-chirality basis spinor
fields \cite{penrose}, \cite{vdw}: 
\begin{equation}
\begin{array}{c}
q_{\alpha }\left( x\right) =\sigma _{\alpha }^{\;A\dot{B}}\ell _{A}\left(
x\right) \otimes r_{\dot{B}}^{\;T}\left( x\right) \text{,\qquad }q^{\alpha
}\left( x\right) =\sigma _{\;\dot{B}A}^{\alpha }r^{\dot{B}}\left( x\right)
^{T}\otimes \ell ^{A}\left( x\right)  \\ 
\bar{q}_{\alpha }=\sigma _{\alpha }^{\;\dot{U}V}r_{\dot{U}}\left( x\right)
\otimes \ell _{V}^{\;T}\left( x\right) \text{,\qquad }\bar{q}^{\alpha
}\left( x\right) =\sigma _{\;V\dot{U}}^{\alpha }\ell ^{V}\left( x\right)
^{T}\otimes r^{\dot{U}}\left( x\right) \text{.}
\end{array}
\label{10}
\end{equation}
We consider the $\mathbf{8}$ basis spinors 
\begin{equation}
\left( \ell _{1},\ell _{2}\right) \text{, }\left( r^{\dot{1}},r^{\dot{2}%
}\right) \text{, }\left( \ell ^{1},\ell ^{2}\right) \text{, }\left( r_{\dot{1%
}},r_{\dot{2}}\right)   \label{11}
\end{equation}
as the \emph{fundamental} physical fields---the \emph{inertial spinor fields}
(I.S.). Dyads in the I.S. are the null tetrads, whose sums and differences
make the (spin-1) spacetime tetrads. Products of these make the (spin-2)
metric tensor.

In fact \cite{penrose}, \cite{vdw}, \cite{spingeo}, all matter and gauge
fields are \emph{spin tensors:} sums of tensor products of spinors of \emph{%
Left} and \emph{Right-chirality} (handed-ness)\emph{. }Outward and Inward 
\emph{temporality} (dilation behavior), and Positive and Negative \emph{%
charge} (temporal $U\left( 1\right) $ current). This is the \cite{penrose}, 
\cite{vdw}, \cite{spingeo}

\begin{quote}
\emph{Spin Principle, }$\mathbf{S}$\emph{:} \emph{Spinor fields} are \emph{%
physical}. All our observable covariants and invariants are sums of tensor
products of spinor fields, and their gradient spinors.
\end{quote}

Immediate implications of $\mathbf{S}$ are:

\begin{enumerate}
\item[A)]  Our moving spacetime tetrads $e_{\alpha }\left( x\right) $ \emph{%
are} the inverse images under (\ref{8}) of \emph{physical} increments $%
q_{\alpha }\left( x\right) $ and $\bar{q}_{\alpha }\left( x\right) $ in spin
space.

\item[B)]  Spinorization maps $S$ and $\bar{S}$ are implemented physically
by the \emph{spin connections} 
\begin{equation*}
\begin{array}{c}
\Omega _{L}\left( x\right) =g^{R}\mathbf{d}g_{L}=\frac{i}{2}\left[
a_{\#}^{-1}q_{\alpha }\left( x\right) +W_{\alpha }\left( x\right) \right]
e^{\alpha } \\ 
\Omega _{R}\left( x\right) =g^{L}\mathbf{d}g_{R}=\frac{i}{2}\left[
a_{\#}^{-1}\bar{q}_{\alpha }\left( x\right) +\bar{W}_{\alpha }\left(
x\right) \right] e^{\alpha }\text{.}
\end{array}
\end{equation*}
The fundamental length unit $a_{\#}$ is the \emph{equilibrium radius} of the
Friedmann solution \cite{im}. In the $PT$-symmetric ($PT_{S}$) case $%
g^{R}g_{L}=\mathbf{1}_{L}^{R}$, the electroweak vector potentials $W_{\alpha
}$ and $\bar{W}_{\alpha }$ vanish. Then the counterpropagating spin-waves in
(\ref{10}) step off our spacetime increments. The spin connections \emph{are}
the tetrads!
\end{enumerate}

For the stationery case $\mathbb{M}_{\#}\equiv \mathbb{S}_{1}\times \mathbb{S%
}_{3}\left( a_{\#}\right) $, the spin connections are the left-invariant
Maurer-Cartan $1$ forms that derive from the canonical maps of $\mathbb{M}%
_{\#} $ onto $U\left( 1\right) \times SU\left( 2\right) $, 
\begin{equation}
\begin{array}{c}
g_{L}=\exp \frac{i}{2a_{\#}}x^{\alpha }\sigma _{\alpha }\text{;\qquad }g_{R}=%
\bar{g}_{L}\text{;} \\ 
\hat{\Omega}_{L}=\frac{i}{2a_{\#}}\sigma _{\alpha }e^{\alpha } \\ 
\hat{\Omega}_{R}=\frac{i}{2a_{\#}}\bar{\sigma}_{\alpha }e^{\alpha }\text{.}
\end{array}
\label{12}
\end{equation}
Their right-invariant versions are 
\begin{equation}
\begin{array}{c}
\Omega ^{L}\equiv \left( \mathbf{d}g^{L}\right) g_{R}\text{;} \\ 
\Omega ^{R}\equiv \left( \mathbf{d}g^{R}\right) g_{L}\text{.}
\end{array}
\label{13}
\end{equation}

It takes tensor products of \emph{all }$4$ chiral pairs of spinor fields in (%
\ref{11}), or their gradients, to make a \emph{natural} $4$ form---e.g. a
Lagrangian density: 
\begin{equation}
\mathcal{L}\in \Lambda ^{4}\subset \otimes ^{8}\text{,}  \label{14}
\end{equation}
which must be invariant under the group $E_{P}$ of passive spin isometries
in curved spacetime. The wedge product of all $4$ spin connections makes the 
$4$\emph{-volume} form $d^{4}V$: 
\begin{equation*}
\begin{array}{c}
i\Omega ^{L}\wedge \Omega _{L}\wedge \Omega ^{R}\wedge \Omega _{R} \\ 
=\left( \frac{1}{16a_{\#}^{4}}\right) \left| g\right| ^{\frac{1}{2}}\sigma
_{0}e^{0}\wedge e^{1}\wedge e^{2}\wedge e^{3}\text{,}
\end{array}
\end{equation*}
where $\left| g\right| $ is the determinant of the metric tensor (\ref{9}).

The \emph{simplest} Lagrangian that is an $E_{P}$-invariant (conformal) $4$
form is the \emph{topological Lagrangian} 
\begin{equation}
\mathcal{L}_{T}\equiv \frac{i}{2}Tr\Omega ^{L}\wedge \Omega _{R}\wedge
\Omega ^{R}\wedge \Omega _{L}\text{,}  \label{15}
\end{equation}
the \emph{Maurer-Cartan} $4$ form. Its action 
\begin{equation}
S_{T}\equiv \frac{i}{2}\int_{\mathbb{M}}Tr\Omega ^{L}\wedge \Omega
_{R}\wedge \Omega ^{R}\wedge \Omega _{L}=-16\pi ^{3}N  \label{16}
\end{equation}
measures the \emph{covering number} of spin space over the ``vacuum'' $%
\mathbb{M}\equiv \mathbb{M}_{\#}\backslash \cup D_{J}$ outside the
worldtubes $D_{J}$ of massive particles, and comes in
topologically-quantized units \cite{tqac}.

Masses arise from the breaking of scale invariance. Elsewhere \cite{capp}, 
\cite{spingeo} we exhibit a ``grandparent'' Lagrangian density for both the 
\emph{outer} region, where it reduces to $\mathcal{L}_{T}$, and the inner
region, where it gives the Dirac Lagrangian, $\mathcal{L}_{D}$: 
\begin{equation}
\mathcal{L}_{G}=i\mathbf{d}\psi ^{R_{\pm }}\psi _{L_{\mp }}\wedge \psi
^{L_{\pm }}\mathbf{d}\psi _{R_{\mp }}\wedge \mathbf{d}\psi ^{L_{\pm }}\psi
_{R_{\mp }}\wedge \psi ^{R_{\pm }}\mathbf{d}\psi _{L_{\mp }}  \label{17}
\end{equation}
(average over sign combinations in which each spinor field appears exactly 
\emph{once}). This is the $\mathbf{8}$\emph{-spinor factorization} of the 
\emph{Maurer-Cartan }$4$\emph{\ form}. Each field is expanded as the sum of
its vacuum distribution and ``broken out'' perturbation 
\begin{equation*}
\begin{array}{ccc}
\psi _{L_{\pm }}\equiv \hat{\psi}_{L_{\pm }}+\tilde{\psi}_{L_{\pm }}\text{,}
&  & \psi _{R_{\pm }}\equiv \hat{\psi}_{R_{\pm }}+\tilde{\psi}_{R_{\pm }}%
\text{,} \\ 
\psi ^{L_{\pm }}\equiv \hat{\psi}^{L_{\pm }}+\tilde{\psi}^{L_{\pm }}\text{,}
&  & \psi ^{R_{\pm }}\equiv \hat{\psi}^{R_{\pm }}+\tilde{\psi}^{R_{\pm }}%
\text{,}
\end{array}
\end{equation*}
where all $\mathbf{8}$ fields may be varied independently.

The outer solution turns out to be $PTC$-symmetric, with $\psi ^{R_{\pm
}}\psi _{L_{\mp }}=1$ \cite{spingeo}. Inside some worldtubes $B_{4}$, chiral
pairs of broken-out perturbations can \emph{bind} to form localized $PT_{A}$%
, or \emph{charged}, \emph{bispinor particles} like 
\begin{equation}
e_{-}\equiv \left( \tilde{\xi}_{-}\left( x\right) \oplus \tilde{\eta}%
_{-}\left( x\right) \right) \text{,}  \label{18}
\end{equation}
which we identify as an \emph{electron.}

The mechanism that endows such bispinors with \emph{inertial mass} emerges
in a remarkable way \cite{im} from Lagrangian (\ref{17}) \cite{im}, \cite
{spingeo}. Inside the worldtube $B_{4}$, $\tilde{\xi}_{-}$ and $\tilde{\eta}%
_{-}$ undergo \emph{mass scatterings }\cite{penrose}, nonlinear resonances
with ``vacuum gratings'' formed by the remaining $3$ \emph{unbroken} chiral
pairs; e.g. on $\mathbb{M}_{\#}$, 
\begin{equation}
\begin{array}{c}
\hat{\Omega}^{3}\equiv \hat{\Omega}_{1}\wedge \hat{\Omega}_{2}\wedge \hat{%
\Omega}_{3} \\ 
=\left( \frac{i}{2a_{\#}}\right) ^{3}\sigma _{1}\sigma _{2}\sigma
_{3}e^{1}\wedge e^{2}\wedge e^{3} \\ 
=\left( \frac{1}{2a_{\#}}\right) ^{3}\gamma ^{-1}\sigma _{0}d^{3}v\text{.}
\end{array}
\label{19}
\end{equation}
The three vacuum spin connections $\hat{\Omega}^{3}$ of (\ref{19})
reconstruct the \emph{spatial }$3$\emph{-volume element,} up to a scale
factor $\gamma ^{-1}$, when inserted into Lagrangian density (\ref{17}).
Here 
\begin{equation}
\gamma \left( T\right) \equiv \frac{a\left( T\right) }{a_{\#}}
\end{equation}
is the time-dependent radius of our Friedmann universe, in units of $a_{\#}$.

Now here is the remarkable thing that happens. When pairs of ``envelope
modulations'' like (\ref{18}) are inserted into Lagrangian (\ref{17}), along
with the three vacuum fields $\hat{\Omega}^{3}$ of (\ref{19}), we obtain an
effective \emph{Dirac} Lagrangian $\mathcal{L}_{D}$ \cite{im}, coupling $%
\tilde{\xi}_{-}\left( x\right) $ and $\tilde{\eta}_{-}\left( x\right) $ via
the effective mass term 
\begin{equation}
\frac{1}{2a_{\#}}\left[ \tilde{\xi}^{-}\tilde{\eta}_{-}-\tilde{\eta}^{-}%
\tilde{\xi}_{-}\right] \text{,}  \label{20}
\end{equation}
where 
\begin{equation}
\tilde{\xi}^{-}=\xi _{-}^{T}iq^{2}\equiv \left( \tilde{\xi}_{-}\right) ^{T}%
\left[ \ell ^{1}\otimes r^{\dot{2}}-\ell ^{2}\otimes r^{\dot{1}}\right]
\equiv \left( \tilde{\xi}_{-}\right) ^{T}\gamma \epsilon   \label{21}
\end{equation}
is the \emph{conformal dual} spinor to $\tilde{\xi}_{-}$. Note that it is
the product (\ref{21}) of two \emph{vacuum fields} that ``dualizes'' each
perturbed envelope to create the \emph{Dirac mass} term, (\ref{20}). This is
Mach's principle in action. $\mathcal{L}_{D}$ gives the \emph{Dirac equations%
} 
\begin{equation}
\begin{array}{c}
i\sigma ^{\alpha }\partial _{\alpha }\tilde{\xi}_{-}=\frac{\gamma ^{-1}}{%
2a_{\#}}\tilde{\eta}_{-} \\ 
i\bar{\sigma}^{\alpha }\partial _{\alpha }\tilde{\eta}_{-}=\frac{\gamma ^{-1}%
}{2a_{\#}}\tilde{\xi}_{-}\text{,}
\end{array}
\label{23}
\end{equation}
written with respect to intrinsic coordinates on our expanding Friedmann $3$%
-brane.

On a microscopic scale, ``mass scatterings'' off vacuum gratings (\ref{19})
are what channel the ``null zig-zags'' of the Dirac propagator \cite{xueg}, 
\cite{penrose}, \cite{ord} into a timelike worldtube $B_{4}\left( \tau
\right) $. The electron mass \cite{im} 
\begin{equation}
m_{e}=\frac{\gamma ^{-1}}{2a_{\#}}  \label{24}
\end{equation}
turns out to be (half) the inverse of the equilibrium radius $a_{\#}$, in
the $\mathbb{M}_{\#}$ reference frame. This must be divided by the scale
factor, $\gamma $, to get the mass we measure in our dilated, intrinsic
frame on $M$.

\section{Matching Boundary Vorticity $\ast G$ to Energy Flux $\ast T$}

The form of energy-momentum flux \emph{inside} $B_{4}$ depends on the
particle. But, the form of the \emph{field} Lagrangian $\mathcal{L}_{G}$
outside the worldtubes $B_{4}$ is universal. This gives us enough
information to match the integrals of the \emph{outer\ field} $3$ form $\ast
G$ and \emph{inner} \emph{matter} flux *T on the \emph{moving} \emph{%
boundaries} $B_{3}\left( \tau \right) =\partial B_{4}\left( \tau \right) $,
and thus \emph{derive} Einstein's field equations. The steps are these:

\begin{enumerate}
\item  Write the total action $S_{G}$ as the sum of \emph{field} terms \emph{%
outside} $B_{4}$ (in $\mathbb{M}_{\#}\backslash B_{4}$), and \emph{matter}
terms \emph{inside}: 
\begin{equation*}
S_{G}=i\int_{\mathbb{M}}Tr\Omega ^{L}\wedge \Omega _{L}\wedge \Omega
^{R}\wedge \Omega _{R}+\int_{B_{4}\left( \tau \right) }\mathcal{L}_{M}\text{.%
}
\end{equation*}
where $\tau $ is a proper time parameter along the particles' world tubes.

\item  Transform the \emph{field} term via integration by parts using the
Bianchi identity 
\begin{equation*}
\mathbf{d}K=K\wedge \Omega -\Omega \wedge K\text{.}
\end{equation*}
The result is 
\begin{equation}
\begin{array}{c}
S_{G}=i\int_{\mathbb{M}}TrK_{L}\wedge K_{R}+Tr\Omega _{L}\wedge \left(
K_{L}+K_{R}\right) \wedge \Omega _{R} \\ 
-i\int_{\partial B_{4}\left( \tau \right) }Tr\left[ \Omega _{L}\wedge
K_{R}-K_{L}\wedge \Omega _{R}\right] \text{.}
\end{array}
\label{25}
\end{equation}
The first and second terms are chiral versions of the electroweak\TEXTsymbol{%
\backslash}strong and gravitational field actions \cite{xueg}, \cite{spingeo}%
.
\end{enumerate}

But it is the third term---the \emph{boundary integral}---that couples
fields to source currents in the next steps.

\begin{enumerate}
\item[3.]  Rewrite the boundary integral in terms of the matrix-valued \emph{%
spacetime curvature} $2$ form \cite{mtw} 
\begin{equation}
\mathcal{R}_{\alpha }^{\;\beta }\equiv R_{\alpha \;\gamma \delta }^{\;\beta
}e^{\gamma }\wedge e^{\delta }\text{.}  \label{26}
\end{equation}
$\mathcal{R}$ accepts an area element and returns the holonomy \emph{%
(rotation)} matrix around it, with matrix elements $\mathcal{R}_{\alpha
}^{\;\beta }$.

\item[4.]  Rewrite all spacetime vectors as Clifford ($C$) vectors, using
spinorization maps (\ref{8}). Now re-express the $PT_{S}$ spacetime
curvature matrix on a basis $C$ vector in terms of $C$ vectors multiplying
the $C$ vector-valued \emph{spin-curvature} $2$ forms 
\begin{equation}
-\mathcal{R}_{\alpha }^{\;\beta }q_{\beta }=q_{\alpha }K_{R}-K_{L}\bar{q}%
_{\alpha }\text{.}  \label{27}
\end{equation}

\item[5.]  Using Cartan's $C$ vector-valued $1$ form 
\begin{equation}
\mathbf{d}q\left( x\right) \equiv \mathbf{d}\left( q_{\alpha }x^{\alpha
}\right) \equiv q_{\alpha }e^{\alpha }\text{,}  \label{28}
\end{equation}
recognize Wheeler's \cite{mtw} ``moment of rotation tensor'' as the $C$
vector-valued $3$ form 
\begin{equation}
\ast G\equiv \mathbf{d}q\wedge \mathcal{R}=2ia_{\#}\left[ \Omega _{L}\wedge
K_{R}-K_{L}\wedge \Omega _{R}\right] \text{.}  \label{29}
\end{equation}

This is the \emph{integrand} in the \emph{outer form} of the boundary
integral in (\ref{25})!

\item[6.]  The \emph{inner} form of the boundary integral is the
energy-momentum $\mathcal{P}^{\alpha }$ inside the worldtube $B_{4}$ of a
moving particle. Detect this by displacing $B_{4}$ by $t=\bigtriangleup
x^{\alpha }$ and rewriting the change in the action as a \emph{surface}
integral of the $3$ form \emph{flux} $\ast T^{\alpha }$ across the moving
boundary $\partial B_{4}\left( t\right) $: 
\begin{equation}
\mathcal{P}^{\alpha }\left( t\right) \equiv \int_{\partial B_{4}\left(
t\right) }\ast T^{\alpha }\text{.}  \label{30}
\end{equation}
Here 
\begin{equation}
\ast T^{\alpha }\equiv \left[ \left( \frac{\partial \mathcal{L}}{\partial
\left( \partial _{\alpha }\psi _{I}\right) }\right) \partial _{\beta }\psi
_{I}-\delta _{\beta }^{\alpha }\mathcal{L}\right] \ast e^{\beta }  \label{31}
\end{equation}
is the \emph{energy-momentum} \emph{density}---the Noether current under 
\emph{active} translation in the $e_{\alpha }$ direction. Taking $t=\tau $,
the proper time along a particles worldline, (\ref{30}) gives $\mathcal{P}%
_{0}\left( \tau \right) $, the \emph{energy} contained in the particle's
support $B_{3}\left( \tau \right) $, i.e. its \emph{rest mass}. For the
Dirac Lagrangian $\mathcal{L}_{D}$, we recover the \emph{frequency} from (%
\ref{31}), 
\begin{equation}
\mathcal{P}_{0}\left( \tau \right) \sim \int_{B_{3}\left( \tau \right)
}\left( \frac{\partial \theta ^{0}}{\partial x^{0}}\right) \left( \tau ,%
\mathbf{x}\right) e^{1}\wedge e^{2}\wedge e^{3}.  \label{32}
\end{equation}

\item[7.]  Finally, equate the inner and outer expressions on the moving
boundary 
\begin{equation*}
\partial B_{4}\left( t\right) \equiv B_{3}\left( t\right) -B_{3}\left(
0\right) +S_{2}\times I\left( t\right) 
\end{equation*}
of the worldtube of a particle in an external field to obtain: 
\begin{equation}
\int_{B_{3}\left( t\right) }\ast G=4a_{\#}^{2}\int_{B_{3}\left( t\right)
}\ast T\Longrightarrow \ast G=4a_{\#}^{2}\ast T\text{,}  \label{33}
\end{equation}
since both integrals must be Lorenz-covariant. These are Einstein's field
equations \cite{mtw}, with a gravitational constant of 
\begin{equation}
\kappa =\frac{a_{\#}^{2}}{2\pi ^{2}}\text{.}  \label{34}
\end{equation}
\end{enumerate}

Using a \emph{spinfluid model}, we have seen how vorticity $\ast G$ arises
on the boundary of each energy-momentum current $\ast T$. Neighboring
centrifugal currents (masses) present \emph{opposite} radiotemporal
vorticities $G_{or}$ to each other's worldtube boundaries---and therefore 
\emph{attract} (or \emph{advect}, like hydrodynamic vortices \cite{neu}).
Spinfluid models give a \emph{mechanism} for gravitation.

The power of such theories to determine some ``constants of nature'' also
makes them \emph{falsifiable. }For example, relations (\ref{24}) and (\ref
{34}) give the value 
\begin{equation}
\kappa m_{e}^{2}=\frac{\gamma ^{-2}}{8\pi ^{2}}\left( T\right)  \label{35}
\end{equation}
for the dimensionless constant that measures the ratio of gravitational to
electromagnetic forces between electrons on our Friedmann $3$-brane, $%
S_{3}\left( a\left( T\right) \right) $. This would match the value observed
today with an expansion factor of $\gamma \sim 10^{20}$.

\section{Quantum Gravity from Null Zig-Zags}

Recall \cite{grosse} that quantum field theory is statistical mechanics in
imaginary time. Cosmic time $T\equiv y^{0}$ enters kinematically as the 
\emph{imaginary} part of a complex time variable $z^{0}\equiv x^{0}+iy^{0}$
in spinfluid models. A stochastic version of our model in which the
classical action is replaced by a statistical propagator---the sum over null
zig-zags---gives a theory of quantum gravity, \emph{provided} that the \emph{%
vacuum fields} that do the mass scatterings are also modelled statistically.

The advantage of such \emph{quantum spinfluid} models is that there are no
divergences built in, so we don't have to worry about unbounded vacuum
energies rolling up our space to a point. The vacuum energy---or \emph{dark
energy}---is simply the energy of the homogeneous distribution of spinor
fields $\left\{ \hat{\psi}_{I},\hat{\psi}^{I}\right\} $ on which the matter
gauge fields ride like waves in the ocean.

Microscopically, the chiral pairs of matter fields that break away from this
geometrical-optics flow resolve into a lightlike mesh of null zig-zags,
confined to timelike worldtubes $B_{4}$ \cite{spingeo}, \cite{tqac}. These
may propagate ``forward'' (centrifugally outward, i.e. with the direction of
cosmic expansion) or ``backward'' (inward).

Spinor fields are \emph{lightlike:} their phases $\zeta ^{\alpha }\left(
z\right) $ or $\zeta ^{\alpha }\left( \bar{z}\right) $ may propagate \emph{%
only} along segments of forward characteristics $\gamma _{+}$ or backward
characteristics $\gamma _{-}$\emph{.} The propagator for a chiral bispinor
particle, a massive Dirac wavefunction \cite{penrose}, \cite{ord}, must
therefore be a sum over \emph{null zig-zags} of $L$-chirality \emph{zigs}
and $R$-chirality \emph{zags; }forward and backward envelopes $\left( \tilde{%
\chi}^{+},\tilde{\zeta}^{+}\right) $ and $\left( \tilde{\xi}_{-},\tilde{\eta}%
_{-}\right) $ with \emph{mass scatterings}---nonlinear resonances with the
remaining 4 vacuum fields---at each corner.

Now suppose our expanding spatial $3$-brane $S_{3}\left( T\right) $ passes
through a vertex where a forward zig is scattered into a backward zag, by a
tensor product nonlinearity \cite{im}, \cite{xueg}. In our spacetime slice,
we see the spin-1 component 
\begin{equation*}
\gamma ^{\circlearrowleft }=\tilde{\xi}_{-}\otimes \tilde{\zeta}^{+}\text{,} 
\end{equation*}
a ($r$-helicity) photon.

The 4 spinor fields whose phases $\zeta ^{\alpha }\equiv \theta ^{\alpha
}-i\varphi ^{\alpha }$ propagate along \emph{forward} characteristics 
\begin{equation}
\gamma _{+}\left( \tau \right) :dy^{0}=+dx^{0}=d\tau  \label{36}
\end{equation}
are called \emph{analytic:} 
\begin{equation}
\begin{array}{c}
\frac{\partial \zeta ^{\alpha }}{\partial \bar{z}^{\beta }}=0\Longrightarrow 
\frac{\partial \varphi ^{\alpha }}{\partial y^{\beta }}=-\frac{\partial
\theta ^{\alpha }}{\partial x^{\beta }}\text{;} \\ 
\frac{\partial \theta ^{\alpha }}{\partial y^{\beta }}=+1\frac{\partial
\varphi ^{\alpha }}{\partial x^{\beta }}=\left[ \frac{\mathbf{d}y^{\gamma }}{%
\mathbf{d}x^{\beta }}\right] \frac{\partial \varphi ^{\alpha }}{\partial
y^{\gamma }}\text{;}
\end{array}
\label{37}
\end{equation}
the Cauchy-Riemann (C.R.) equations in $z^{\beta }$. The remaining 4 spinor
fields are \emph{conjugate analytic,} i.e. we must replace $\bar{z}^{\beta }$
with $z^{\beta }$ in (\ref{37}), to get $4$ conjugate C.R. equations.

The C.R. and conjugate C.R. equations relate the $u\left( 1\right) \times
su\left( 2\right) $ phases $\theta ^{\alpha }$ and their complexifications $%
\varphi ^{\alpha }$, given in terms of polar coordinates $\left(
x^{0},x^{j}\right) $ on $\mathbb{M}_{\#}\equiv \mathbb{S}_{1}\times \mathbb{S%
}_{3}$ to their conjugate momenta, $\left( y^{0},y^{j}\right) $. These
analyticity conditions justify \emph{Wick rotation,} which translates the
statistical mechanics of null zig-zags in \emph{Euclidean spacetime} $\left(
y^{0},\mathbf{x}\right) \in \mathbb{M}$ to Feynman integrals in
(compactified) Minkowsky space $\left( x^{0},\mathbf{x}\right) \in \mathbb{M}%
_{\#}$.

These mass scatterings are the vertices in a Riemann sum for the action, $%
S_{G}$, for a massive particle. This becomes clear when we

\begin{enumerate}
\item[i)]  re-express $S_{G}$ with respect to \emph{null tetrads} on $%
\mathbb{M}$, 
\begin{equation}
\begin{array}{c}
e^{\pm }=\frac{1}{\sqrt{2}}\left( e^{1}\pm ie^{2}\right)  \\ 
e^{\uparrow \downarrow }=\frac{1}{\sqrt{2}}\left( e^{0}\pm e^{3}\right) 
\text{.}
\end{array}
\label{38}
\end{equation}

\item[ii)]  write Riemann sums for $S_{G}$ with respect to a null lattice
(e.g. with spacing $a_{\#}$ on $\mathbb{M}$), $N$, stepped off by these.

\item[iii)]  notice that, in order for a lattice point to contribute to the
action, the Lagrangian there must have a scalar $\left( \sigma _{0}\right) $%
, or spin-0 component. We call this a \emph{nonlinear }$\mathbf{8}$\emph{%
-spinor resonance} between $J$ chiral pairs of broken-out matter fields and $%
\left( 4-J\right) $ vacuum pairs. The action in the N.M. model is a sum of
contributions from every such resonance in the null lattice, and the
propagator is the sum over all null zig-zag paths that connect the initial
and final point. In this sense, this N.M. model is innately quantum
mechanical, and does not need to be ``quantized.''
\end{enumerate}

Furthermore, it is the minimal model with a $1$-term, passive-Einstein ($%
E_{P}$)-invariant topological Lagrangian, because:

\begin{enumerate}
\item[a)]  It takes the intersection of $4$ null cones to determine a point
on $\mathbb{M}$.

\item[b)]  Each is generated, via $S^{-1}$ of (\ref{8}), by the tensor
product (\ref{10}) of $2$ spinor fields.

\item[c)]  For symplectic invariance, $\mathcal{L}$ must contain $4$ spinors
and $4$ gradients.

\item[d)]  Under $PTC$, $\mathcal{L}$ reduces to the Maurer-Cartan (M.C.) $4$
form, which measures the covering number of $U\left( 1\right) \times
SU\left( 2\right) $ over our (compactified) spacetime manifold.
\end{enumerate}

I don't know how our N.M. model compares with other theories of quantum
gravity, or what experimental tests could distinguish them at this era of
cosmic expansion. However, the N.M. model is quite capable of dealing with
the \emph{superdense} regimes inside collapsed objects like neutron stars,
black holes, and the early universe, where it admits \emph{nonperturbative}
solutions \cite{im}, \cite{tqac}. These are regularized by the resistance (%
\ref{16}) of the topologically nontrivial vacuum to compression to a point.
The N.M. model predicts another new effect in \emph{supervorticial} regimes:
an interaction between Lens-Thirring fields and weak potentials \cite{xueg}, 
\cite{spingeo}. Perhaps such effects could be measured in terrestrial
laboratories or astronomical observation.

\section{Conclusion}

More significant than the values of fundamental constants derived from the
N.M. model, or the prediction of new effects, are the qualitative features
of \emph{quantum} \emph{spinfluid} models that enable them to reconcile
quantum mechanics and general relativity. These are

\begin{enumerate}
\item  A Lagrangian density with \emph{no} free parameters that is a \emph{%
natural} $4$ form---i.e. invariant under the group of passive spin
isometries in curved spacetime.

\item  An action which \emph{includes} a bounded vacuum energy that depends
on the radius $a\left( T\right) $ of the Friedmann solution which breaks
dilation invariance and sets the length and the mass scales. This
(hopefully) includes a repulsive term at high densities that prevents total
collapse.

\item  Values for the standard coupling constants that are ``frozen in'' by
the history of dynamical symmetry breaking. These values may depend on
cosmic time, $T$.

\item  Effective electroweak, strong, and gravitational \emph{field}
actions---along with minimal coupling through their spin connections in the
covariant derivatives---are \emph{derived} from local perturbations to the
Friedmann vacuum.

\item  These fields are sourced in localized currents with topologically
quantized charges.

\item  A \emph{mechanism} for gravitation derived from the \emph{same}
nonlinear coupling of particle fields to the global field, sourced in the
``distant masses,'' that creates the \emph{inertial masses} of particles.

\item  Quantum propagators which are \emph{derived }from the statistical
mechanics of null zig-zags of the (lightlike) spinor fields that weave the
(timelike) worldtubes of massive particles and the (spacelike) fabric---the 
\emph{vacuum}---that connects them.
\end{enumerate}

\end{document}